\documentclass[aps,pra,twocolumn,showpacs,superscriptaddress,draft,floatfix]{revtex4}
\input{epsf}
\usepackage{amsmath}
\usepackage{amssymb}
\input{epsf}

\newcommand{\rem}[1]{}

\begin{document}

\title{Optimal purification of a generic $n$-qudit state}
\author{Giuliano Benenti}
\email{giuliano.benenti@uninsubria.it}
\affiliation{CNISM, CNR-INFM \& Center for Nonlinear and Complex Systems,
Universit\`a degli Studi dell'Insubria, Via Valleggio 11, 22100 Como, Italy}
\affiliation{Istituto Nazionale di Fisica Nucleare, Sezione di Milano,
via Celoria 16, 20133 Milano, Italy}
\author{Giuliano Strini}
\email{giuliano.strini@mi.infn.it}
\affiliation{Dipartimento di Fisica, Universit\`a degli Studi di Milano,
Via Celoria 16, 20133 Milano, Italy}
%\date{\today}
%\date{October 8, 2008}
\date{January 13, 2009}
\begin{abstract}
We propose a quantum algorithm for the purification of a generic mixed
state $\rho$ of a $n$-qudit system by using an ancillary $n$-qudit system. 
The algorithm is optimal in that (i) the number of ancillary qudits 
cannot be reduced, (ii) the number of parameters which determine
the purification state $|\Psi\rangle$ exactly 
equals the number of degrees of freedom of $\rho$, and  
(iii) $|\Psi\rangle$ is easily determined from the density matrix $\rho$.
Moreover, we introduce a quantum circuit in which the quantum gates 
are unitary transformations acting on a $2n$-qudit system.
These transformations are determined by 
parameters that can be tuned to generate, once the ancillary qudits 
are disregarded, any given mixed $n$-qudit state.
\end{abstract}


% 03.67.-a Quantum information
% 03.67.Ac Quantum algorithms, protocols, and simulations

\maketitle

\section{Introduction}

\label{sec:introduction}

Purification is one of the basic tools in quantum 
information science~\cite{qcbook,nielsen}: Given a 
mixed quantum system $S$ described by a density matrix $\rho$
it is possible to introduce another ancillary system $A$, such that
the state $|\Psi\rangle$ of the composite system $S+A$ is a pure state
and $\rho$ is recovered after partial tracing over $A$:
$\rho={\rm Tr}_A (|\Psi\rangle\langle\Psi|)$. 
The ancillary system may be a physical environment that must be taken
into account when doing experiments on the system $S$, but not
necessarily so. It may be a fictitious environment that allows
us to prove interesting results about the system under investigation.

Purification is a tool of great value in quantum information science,
with countless applications, for instance in the study of the 
distance between quantum states~\cite{nielsen}, of the geometry of 
quantum states~\cite{karol} and of the quantum capacity of noisy
quantum channels~\cite{barnum}. 
Besides its theoretical relevance, purification is interesting 
for experimental implementations of quantum information protocols
requiring mixed state. While the direct generation of a mixture $\rho$
of quantum states necessarily involves statistical errors, this 
problem can be avoided if the purification state $|\Psi\rangle$
is generated. Of course, the price to pay is that one has to work
with the enlarged system $S+A$, including the ancillary system $A$. 

Due to the partial trace structure the purification state $|\Psi\rangle$ 
of a density matrix $\rho$ cannot be uniquely defined, as any unitary 
transformation $U_A\otimes \openone_S$ acting non-trivially on the ancillary 
system only maps the state $|\Psi\rangle$ into a new state
$|\Psi'\rangle=(U_A\otimes \openone_S)|\Psi\rangle $ which is again a 
purification of $\rho$. In this paper, we propose a quantum protocol
that selects a specific purification $|\Psi\rangle$. Such purification
turns out to be very convenient since the state $|\Psi\rangle$ 
depends on a number of parameters exactly equal to the 
number of degrees of freedom of a generic mixed state $\rho$ and is 
easily determined as a function of $\rho$.
Furthermore, we design a quantum circuit given by a sequence of
quantum gates acting on both the system and the ancillary qudits.
The parameters which determine such quantum gates can be tuned to 
generate, once the ancillary qubits are disregarded, any mixed
state $\rho$ of system $S$. Finally, our protocol works for 
any system of $n$ qudits and requires $n$ ancillary qudits.
We show that the number of ancillary qudits is optimal, that is,
it cannot be reduced if we wish to design a quantum circuit 
capable of generating any $n$-qudit mixed state.

Our paper is organized as follows. To set the notations, we first
briefly review the concept of purification (Sec.~\ref{sec:purification}). 
Then we propose our purification scheme, moving from simple to more 
and more complex cases.
We start with the purification of a single qubit mixed state
(Sec.~\ref{sec:qubit}), then we proceed with the qutrit 
(Sec.~\ref{sec:qutrit}) and the two-qubit (Sec.~\ref{sec:2qubits})
cases and finally we illustrate the generic 
$n$-qudit case (Sec.~\ref{sec:nqudits}).
Appendix~\ref{app:dstates} provides a short account of the 
diagrams of states, namely of a tool very useful for the purposes 
of the present paper.  

\section{Purification}

\label{sec:purification}

Given a quantum system $S$ described by the density matrix $\rho$,
it is possible to introduce another ancillary system $A$, such that
the state $|\Psi\rangle$ of the composite system is pure and
\begin{equation}
\rho={\rm Tr}_A (|\Psi\rangle\langle\Psi|).
\label{eq:ptrace}
\end{equation} 
This procedure, known
as purification, allows us to associate a pure state $|\Psi\rangle$ 
with a density matrix $\rho$. 
A generic pure state of the global system $S+A$ is given by 
\begin{equation}
|\Psi\rangle = \sum_{\alpha=0}^{M-1} \sum_{i=0}^{N-1} 
C_{\alpha i} |\alpha\rangle |i\rangle,
\label{eq:psipurification}
\end{equation}
with $\{|\alpha \rangle\}$ and $\{|i \rangle\}$ basis sets for the
Hilbert spaces $\mathcal{H}_A$ and $\mathcal{H}_S$, of dimensions 
$M$ and $N$,
associated with the subsystems $A$ and $S$. 
Given a generic density matrix for system $S$,
\begin{equation}
\rho=\sum_{i,j=0}^{N-1} \rho_{ij} |i\rangle \langle j|,
\label{eq:rhos1}
\end{equation}
we say that the state $|\Psi\rangle$ defined by 
Eq.~(\ref{eq:psipurification}) is a purification of $\rho$ if
\begin{equation}
\rho={\rm Tr}_A (|\Psi\rangle\langle\Psi|)=
\sum_{\alpha=0}^{M-1} \sum_{i,j=0}^{N-1}
C_{\alpha i} C_{\alpha j}^\star |i\rangle \langle j|.
\label{eq:rhos2}
\end{equation}
The equality between (\ref{eq:rhos1}) and (\ref{eq:rhos2})
implies
\begin{equation}
\rho_{ij}=\sum_{\alpha=0}^{M-1} C_{\alpha i} C_{\alpha j}^\star.
\label{purecondition}
\end{equation}
It is clear that (\ref{purecondition}) always admits a solution, 
provided the Hilbert space of system $A$ is large enough.
More precisely, it is sufficient to consider a system $A$
whose Hilbert space dimension is the same as that of system $S$.
Indeed, if we express the reduced density matrix
using its diagonal representation,
\begin{equation}
  \rho = \sum_i p_i |i\rangle\langle i |,
  \label{dmatdiag}
\end{equation}
a purification for the density matrix (\ref{dmatdiag}) is
given by
\begin{equation}
  |\Psi\rangle  = \sum_i \sqrt{p_i} \, |i'\rangle |i\rangle,
\label{eq:spectralpur}
\end{equation}
with $\{|i'\rangle\}$ orthonormal basis for $\mathcal{H}_A$. 
This purification procedure requires the diagonalization of 
the density matrix $\rho$ and therefore is in general 
only numerically feasible.

In what follows, we propose a different purification scheme, 
which is optimal in that, for the purification of a 
generic $n$-qudit state, 
the number $m=n$ of qudits of the ancillary system  $A$ cannot be reduced. 
While this is the case also for the well-known purification
(\ref{eq:spectralpur}) based on the spectral decomposition 
(\ref{dmatdiag}), we anticipate that our method readily provides a 
purification state $|\Psi\rangle$ 
that depends on a number of parameters exactly 
equal to the number of degrees of freedom of a generic $\rho$.
Note that, even when the number of ancillary qudits 
is optimal ($m=n$ so that $M=N$), the number $2 N^2-2$ of real 
coefficients $C_{\alpha i}$ determining the purification state 
(\ref{eq:psipurification}) is in general much larger than the number $N^2-1$ 
of real free parameters that must be set to determine a generic 
density matrix of size $N$. Different choices of the coefficients 
$C_{\alpha i}$ are therefore possible. Our choice provides 
a purification state $|\Psi\rangle$ depending on a number $N^2-1$ 
of real parameters exactly equal to the number of real freedoms 
of a generic mixed states $\rho$.
Furthermore, the coefficient 
$C_{\alpha i}$ in (\ref{eq:psipurification}) can be easily determined
from conditions (\ref{purecondition}), in spite of the fact that 
these equations are nonlinear. Finally, we will see in the next
sections that our scheme suggest a very convenient quantum circuit
for the preparation of a generic density matrix $\rho$.
To illustrate the working of our purification method, we discuss 
cases of increasing complexity, from a single qubit state 
to a generic $n$-qudit state.

\section{Qubit}

\label{sec:qubit}

\subsection{Mixed-state purification}

We consider $M=N=2$ and set 
\begin{equation}
\left\{
\begin{array}{l}
%C_{01}\in \mathbb{R}, \,C_{01}\ge 0\\
%C_{10}\in \mathbb{R}, \,C_{10}\ge 0, \,C_{11}=0.
C_{01}\in \mathbb{R}_+, \\
C_{10}\in \mathbb{R}_+, \,C_{11}=0.
\end{array}
\right.
\label{qubitconstraints}
\end{equation}

We first determine $C_{01}$  
from (\ref{purecondition}): 
\begin{equation}
\rho_{11}=|C_{01}|^2+|C_{11}|^2=|C_{01}|^2,
\label{eq:rho11}
\end{equation}
where the last equality follows from the second line of
(\ref{qubitconstraints}).
Since we have also chosen $C_{01}$ to be real and 
nonnegative, we simply obtain
\begin{equation}
C_{01}=\sqrt{\rho_{11}}.
\end{equation}
Then we can determine $C_{00}$ from the condition
\begin{equation}
\rho_{01}=C_{00}C_{01}^\star+C_{10}C_{11}^\star=
C_{00}C_{01},
\end{equation}
since $C_{01}$ is already known from (\ref{eq:rho11}).
Finally, knowing $C_{00}$, we can derive $C_{10}$ from the 
condition
\begin{equation}
\rho_{00}=|C_{00}|^2+|C_{10}|^2.
\end{equation}
Taking into account that $C_{10}\in\mathbb{R}_+$,
we have
\begin{equation}
C_{10}=\sqrt{\rho_{00}-|C_{00}|^2}.
\end{equation}  

Note that, in the special case in which $\rho_{11}=0$ we can remove 
any ambiguity in the definition of the purification state $|\Psi\rangle$
by setting $C_{00}=0$. 
In this case, $\rho=|0\rangle\langle 0|$ is
already a pure state and its ``purification'' 
is $|\Psi\rangle=|10\rangle$.
Alternatively, one can reshuffle the basis 
state according to 
$|0\rangle \leftrightarrow |1\rangle$\footnote{Similar procedures can 
be applied to higher dimensional cases 
whenever diagonal elements of the density matrix $\rho$ are equal to zero.
For the sake of simplicity, we will not discuss any longer in our paper 
such special cases.}. 
For a generic state $\rho_{11}\ne 0$ and the state $|\Psi\rangle$ reads,
in the $\{|\alpha i\rangle = 
|00\rangle,
|01\rangle,
|10\rangle,
|11\rangle\}$ basis, as follows:
\begin{equation}
|\Psi\rangle=
\left[
\begin{array}{c}
\displaystyle C_{00}
\\
\\
\displaystyle C_{01}
\\
\\
\displaystyle C_{10}
\\
\\
\displaystyle C_{11}
\end{array}
\right
]=
\left[
\begin{array}{c}
{\displaystyle
\frac{\rho_{01}}{\sqrt{\rho_{11}}}
}
\\
\\
{\displaystyle
\sqrt{\rho_{11}}
}
\\
\\
{\displaystyle
\sqrt{
\frac{\rho_{00}\rho_{11}-\rho_{10}\rho_{01}}{\rho_{11}}
}
}
\\
\\
0
\end{array}
\right].
\label{eq:Psi1q}
\end{equation}

\subsection{Mixed-state generation}

In this subsection, we provide a quantum circuit generating 
the state $|\Psi\rangle$ of (\ref{eq:Psi1q}), 
namely the purification of a generic single-qubit 
density matrix $\rho$. 
After disregarding the ancillary qubit (this corresponds
to performing the partial trace of the density matrix
$|\Psi\rangle\langle \Psi|$ over the ancillary qubit), 
we obtain the mixed state $\rho$. Therefore, we end up
with an experimentally viable procedure 
for the generation of a generic mixed single-qubit state
by means of a two-qubit state subjected??? to controlled unitary 
transformations. 

\begin{figure}
\centerline{\epsfxsize=8.cm\epsffile{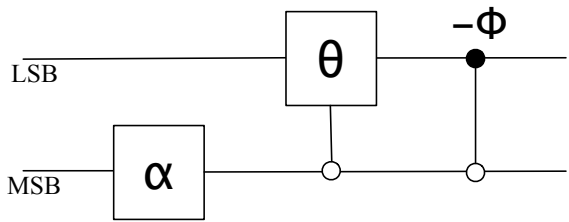}}
\caption{Quantum circuit for the purification of a single qubit.
The qubits run from top to bottom from the least significant (LSB) 
to the most significant (MSB).}
\label{fig:qcircuit1qubit}
\end{figure}

The quantum circuit generating the state $|\Psi\rangle$ 
is shown in Fig.~\ref{fig:qcircuit1qubit}.
A square box with a greek letter inside (here, $\alpha$ or
$\theta$) stands for a 
rotation operator. Its matrix representation in
the $\{|0\rangle,|1\rangle\}$ reads
as follows:
\begin{equation}
R(\alpha) =
\left[
\begin{array}{cc}
\cos\alpha & -\sin\alpha \\
\sin\alpha & \cos\alpha
\end{array}
\right].
\label{eq:Ralpha}
\end{equation}
The full circle with $-\phi$ above is the phase-shift gate,
defined by the diagonal matrix
\begin{equation}
{\rm PHASE}(-\phi)= 
{\rm diag}(1,e^{-i\phi}).
\label{eq:PHASE}
\end{equation}
As an overall phase factor is arbitrary, the action of this
gate is equivalently represented by the matrix
${\rm diag}(e^{i\phi},1)$.
In the controlled-gates, the 
empty circle on the control qubit 
means that the gate acts non trivially (differently from identity)
on the target qubit if and only if the state of the control qubit
is $|0\rangle$.

On the other hand, any single-qubit density matrix can be generated
by means of the quantum circuit in Fig.~\ref{fig:qcircuit1qubit}.
Given the input state $|\Psi_i\rangle=|00\rangle$, the output state is
\begin{equation}
|\Psi_f\rangle =
\left[
\begin{array}{c}
\cos \alpha \cos \theta e^{i\phi} \\
\cos \alpha \sin \theta  \\
\sin \alpha \\
0 
\end{array}
\right].
\label{eq:Psio1q}
\end{equation}
This state is equal to the purification state (\ref{eq:Psi1q}),
provided we set
\begin{equation}
\left\{
\begin{array}{l}
\displaystyle
\sin\alpha=C_{10},\\
\displaystyle
\cos\alpha\sin\theta=C_{01},\\
\displaystyle
\cos\alpha\cos\theta e^{i\phi}=C_{00}.
\end{array}
\right.
\end{equation}
From the first equation we determine $\alpha$, then from the
second $\theta$ and finally from the third $\phi$, as a function of
the coefficients $C_{\alpha i}$, 
which in turn are determined by our purification
protocol from the density matrix elements $\rho_{ij}$.  
Note that, since by construction $C_{01}, C_{10}\ge 0$, we can take 
$\alpha,\theta\in\left[0,\frac{\pi}{2}\right]$. 
Finally, the phase $\phi\in [0,2\pi)$. 

\begin{figure}
\centerline{\epsfxsize=8.cm\epsffile{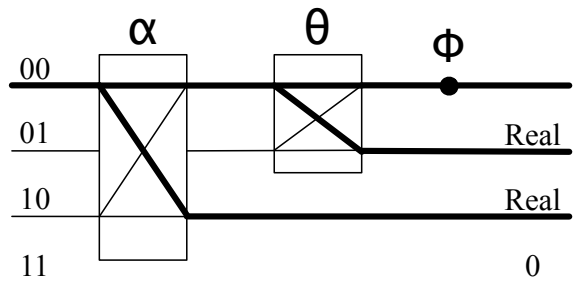}}
\caption{Diagram of states for the purification of a single qubit.
Starting from the input state $|00\rangle$, information flows on the 
thick lines. We esplicitly indicate at the right hand side that 
the coefficientes $C_{01}$ and $C_{10}$ are real, while $C_{11}=0$.}
\label{fig:dstates1qubit}
\end{figure}

For a straightforward extension of the results presented in this section
from the purification of a single qubit to more complex systems it is 
convenient to express the quantum circuit in Fig.~\ref{fig:qcircuit1qubit}  
in terms of diagrams of states~\cite{diagrams}, of which a very brief 
account is given in  
Appendix~\ref{app:dstates}. The diagram of states corresponding to
the purification circuit of Fig.~\ref{fig:qcircuit1qubit} is shown in 
Fig.~\ref{fig:dstates1qubit}. Note that, given the input state $|00\rangle$,  
the output state (\ref{eq:Psio1q}) is immediately written 
following the information flow along the thick lines of the 
diagram of states.
We stress that other optimal purifications, where the 
purification state is determined by $N^2-1=3$ real parameters,
are possible. Our choice corresponds to setting $C_{11}=0$ and 
$C_{01}$, $C_{10}$ real. We will see that the diagrams of states 
immediately lead to optimal purifications also for arbitrarily complex
systems. 

\subsection{``Invasion'' of the Bloch ball}

\label{sec:invasion}

\begin{figure}
\centerline{\epsfxsize=4.2cm\epsffile{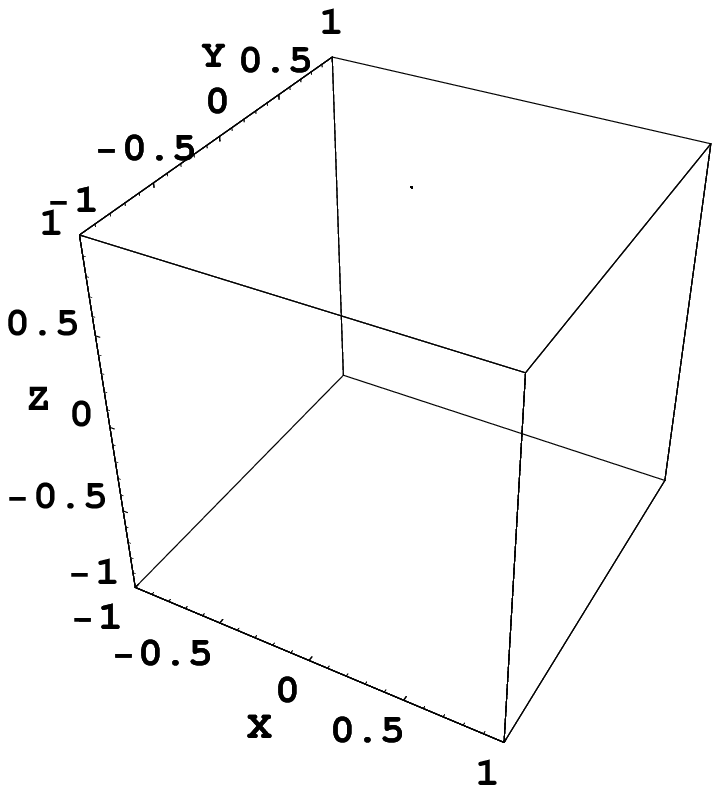}
\hfill\epsfxsize=4.2cm\epsffile{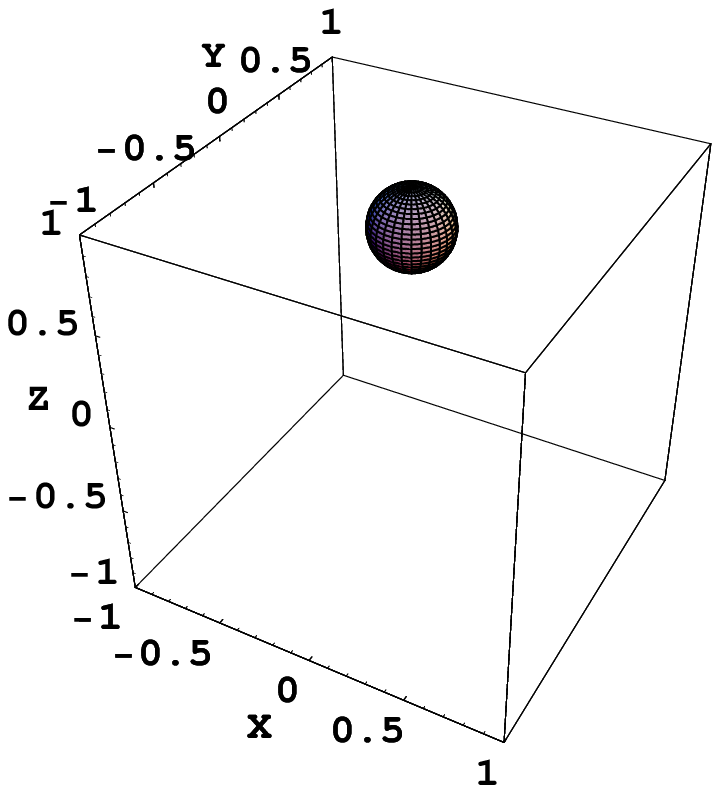}}
%\vspace{-.2cm}
\centerline{\epsfxsize=4.2cm\epsffile{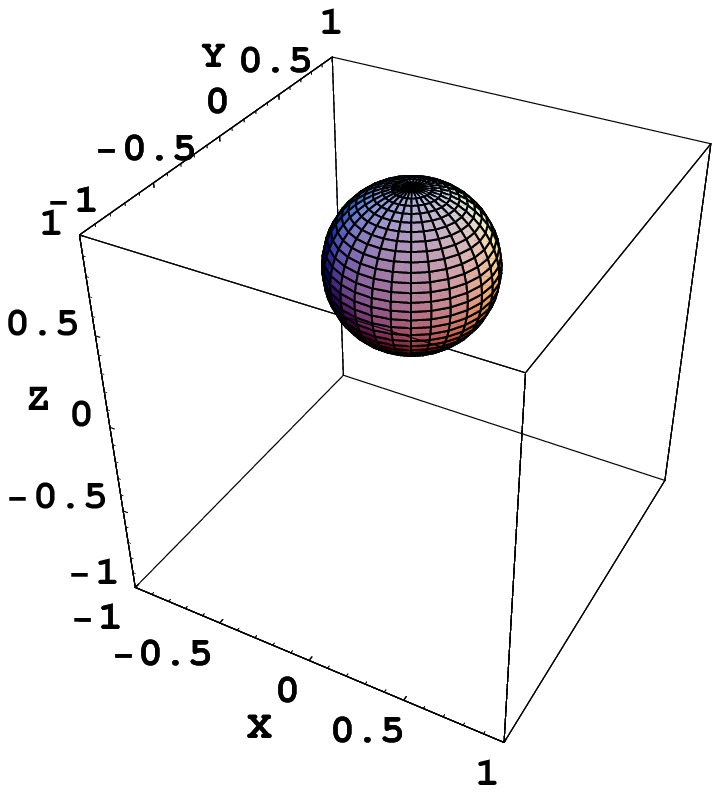}
\hfill\epsfxsize=4.2cm\epsffile{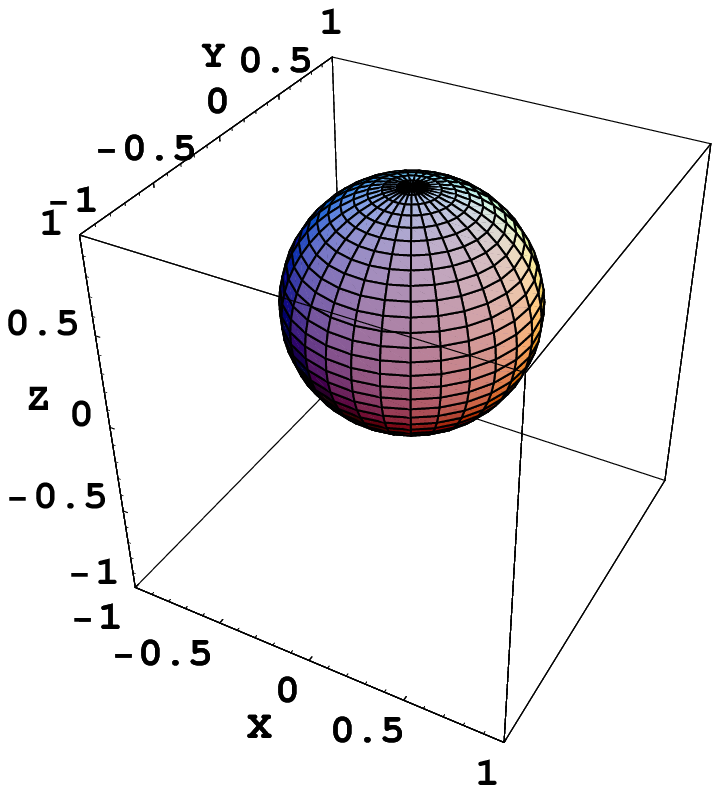}}
%\vspace{-0.2cm}
\centerline{\epsfxsize=4.2cm\epsffile{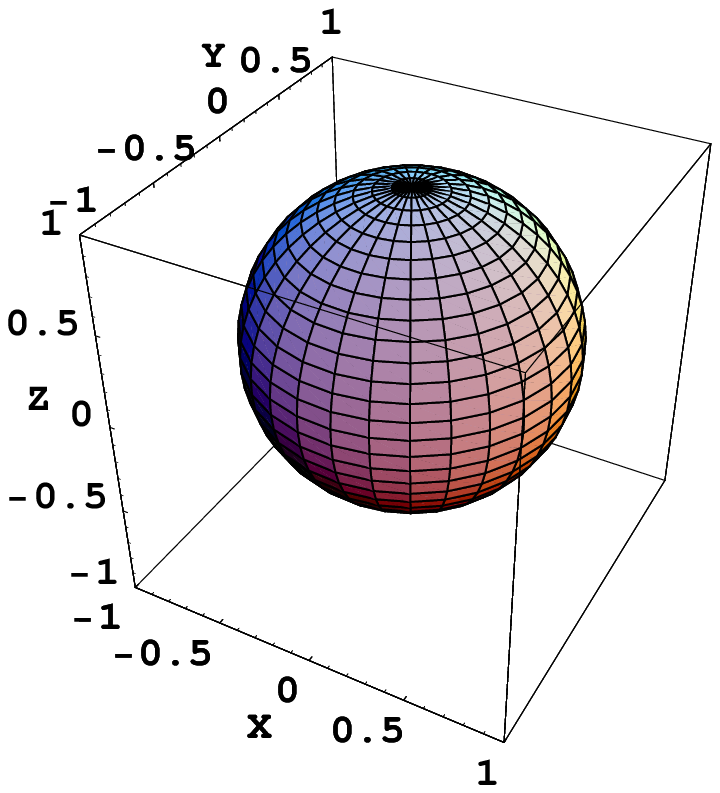}
\hfill\epsfxsize=4.2cm\epsffile{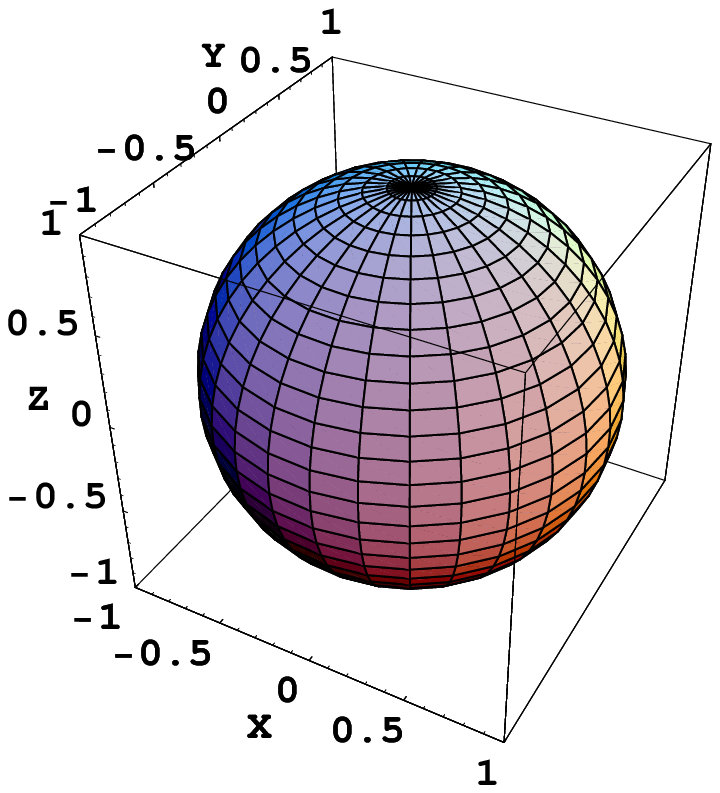}}
\caption{``Invasion'' of the Bloch ball:
$\cos^2\alpha=0$ (top left),
$\frac{\pi}{10}$ (top right),
$2\frac{\pi}{10}$ (middle left),
$3\frac{\pi}{10}$ (middle right),
$4\frac{\pi}{10}$ (bottom left),
$\frac{\pi}{2}$ (bottom right).}
\label{fig:sphere}
\end{figure}

Using the quantum circuit in Fig.~\ref{fig:qcircuit1qubit} we 
can write a generic single-qubit state as 
\begin{equation}
\rho= \sum_{k=1}^2 p_k |\psi_k\rangle \langle \psi_k|,
\end{equation}
with $p_1=\cos^2\alpha$, $p_2=\sin^2 \alpha$,
$|\psi_1\rangle$ generic single-qubit pure state
and $|\psi_2\rangle=|0\rangle$. The matrix representation of 
$\rho$ in the $\{|0\rangle,|1\rangle\}$ basis is given by 
\begin{eqnarray}
{\displaystyle
\rho=\cos^2\alpha
\left[
\begin{array}{cc}
\cos^2 \theta & \cos\theta \sin\theta e^{i\phi} \\
\cos\theta \sin\theta e^{-i\phi} & \sin^2\theta
\end{array}
\right]
\nonumber
}
\\
{\displaystyle
+\sin^2\alpha 
\left[
\begin{array}{cc}
 1 & 0 \\
 0 & 0
\end{array}
\right]=
\frac{1}{2}
\left[
\begin{array}{cc}
 1+Z & X-iY \\
 X+iY & 1-Z
\end{array}
\right].
}
\label{eq:Bloch1q}
\end{eqnarray}
The last equality corresponds to the usual Bloch-ball representation
of the (generally mixed) single-qubit states. 
It is clear that, once $\alpha$ is fixed, 
Eq.~(\ref{eq:Bloch1q}) represents a surface
in the $(X,Y,Z)$-space, obtained after contracting the pure-states
(unit radius) Bloch-sphere of the factor $\cos^2\alpha$ and 
translating it in the positive $Z$-direction by $\sin^2\alpha$.
Plots of this surface for different values of $\alpha$ are shown 
in Fig.~\ref{fig:sphere}. It is clear that all the points of 
the Bloch ball are recovered when $\alpha$ goes from $0$ to $\frac{\pi}{2}$.
We can say that there is an ``invasion'' of the Bloch ball starting
from the north pole.

\section{Qutrit}

\label{sec:qutrit}

We consider $M=N=3$ and set 
\begin{equation}
\left\{
\begin{array}{l}
C_{02}\in \mathbb{R}_+,\\
C_{11}\in \mathbb{R}_+,\, C_{12}=0,\\
C_{20}\in \mathbb{R}_+,\, C_{21}=C_{22}=0.
%C_{02}\in \mathbb{R},\,C_{02}\ge 0, \\
%C_{11}\in \mathbb{R},\,C_{11}\ge 0, \,C_{12}=0,\\
%C_{20}\in \mathbb{R},\,C_{20}\ge 0, \,C_{21}=C_{22}=0.
\end{array}
\right.
\label{eq:qutritgauge}
\end{equation}

We first obtain $C_{02}$ from 
\begin{equation}
\rho_{22}=|C_{02}|^2+|C_{12}|^2+|C_{22}|^2=|C_{02}|^2.
\end{equation}
Once $C_{02}\ge 0$ is determined as 
$C_{02}=\sqrt{\rho_{22}}$, 
we obtain $C_{01}$ and $C_{00}$ from 
\begin{equation}
\rho_{12}=C_{01}C_{02}^\star+C_{11}C_{12}^\star+C_{21}C_{22}^\star=
C_{01}C_{02},
\end{equation}
\begin{equation}
\rho_{02}=C_{00}C_{02}^\star+C_{10}C_{12}^\star+C_{20}C_{22}^\star=
C_{00}C_{02}.
\end{equation}
Then we obtain $C_{11}$ from
\begin{equation}
\rho_{11}=|C_{01}|^2+|C_{11}|^2+|C_{21}|^2=
|C_{01}|^2+C_{11}^2,
\end{equation}
and, finally, $C_{10}$ and $C_{20}$ from
\begin{equation}
\rho_{01}=C_{00}C_{01}^\star+C_{10}C_{11}^\star+C_{20}C_{21}^\star
=C_{00}C_{01}^\star+C_{10}C_{11},
\end{equation}
\begin{equation}
\rho_{00}=|C_{00}|^2+|C_{10}|^2+C_{20}^2.
\end{equation}

\begin{figure}
\centerline{\epsfxsize=8.cm\epsffile{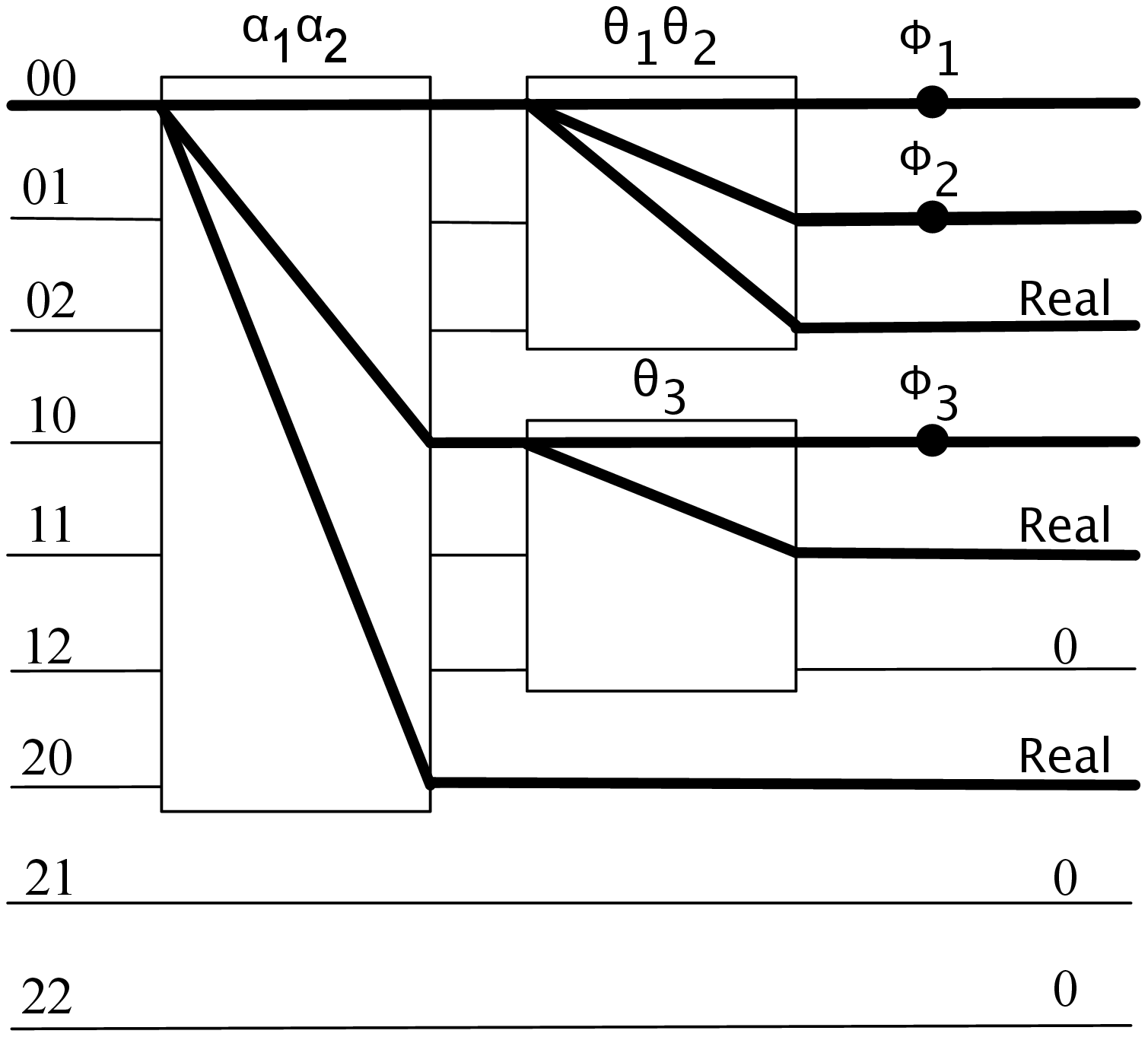}}
\caption{Diagram of states for the purification of a single qutrit.
To simplify the plot, only the thick lines corresponding to the
information flow are shown inside the boxes. The angles 
$\alpha_i,\theta_j\in\left[0,\frac{\pi}{2}\right]$, while the 
phases $\phi_k\in[0,2\pi)$.}
\label{fig:dstates1qutrit}
\end{figure}

We stress that conditions (\ref{eq:qutritgauge}) 
lead to a purification state determined by 
$N^2-1=8$ free real parameters,
exactly corresponding to the number of real freedoms
needed to set a generic density matrix for a qutrit.
Conditions (\ref{eq:qutritgauge}) are readily derived if the 
purification of a generic
qutrit state is implemented by means of the diagram of states 
shown in Fig.~\ref{fig:dstates1qutrit}.
In this figure, a box with two geek letters written on top of it 
(for instance, $\alpha_1$ and $\alpha_2$) represents a
unitary transformation whose matrix representation has, in the 
$\{|0\rangle,|1\rangle,|2\rangle\}$ basis, the first column given by 
\begin{equation}
\left[
\begin{array}{c}
\cos\alpha_1 \cos\alpha_2
\\
\cos\alpha_1 \sin\alpha_2
\\
\sin\alpha_1
\end{array}
\right].
\end{equation}
Such transformation maps the input state $|0\rangle$ into 
\begin{equation}
\cos\alpha_1\cos\alpha_2|0\rangle+
\cos\alpha_1\sin\alpha_2|1\rangle+
\sin\alpha_1|2\rangle.
\end{equation}
Finally, the box with the letter $\theta_3$ on top of it represents 
the rotation $R(\theta_3)$ acting on the two-dimensional subspace
spanned by the state $|0\rangle$ and $|1\rangle$, 
with $R$ defined by (\ref{eq:Ralpha}).

Given the input state $|00\rangle$, the output purification state
$|\Psi\rangle=\sum_{\alpha,i} C_{\alpha i}|\alpha i\rangle$
can be immediately written by following the thick lines of 
the diagram of states in Fig.~\ref{fig:dstates1qutrit}.
We obtain
\begin{equation}
\left\{
\begin{array}{l}
\displaystyle
C_{00}=\cos\alpha_1\cos\alpha_2\cos\theta_1\cos\theta_2e^{i\phi_1},\\
\displaystyle
C_{01}=\cos\alpha_1\cos\alpha_2\cos\theta_1\sin\theta_2e^{i\phi_2},\\
\displaystyle
C_{02}=\cos\alpha_1\cos\alpha_2\sin\theta_1,\\
\displaystyle
C_{10}=\cos\alpha_1\sin\alpha_2\cos\theta_3e^{i\phi_3},\\
\displaystyle
C_{11}=\cos\alpha_1\sin\alpha_2\sin\theta_3,\\
\displaystyle
C_{12}=0,\\
\displaystyle
C_{20}=\sin\alpha_1,\\
\displaystyle
C_{21}=0,\\
\displaystyle
C_{22}=0.
\end{array}
\right.
\end{equation}
These relations can be easily inverted to obtain the parameters
$\{\alpha_j,\theta_k,\phi_l\}$ as a function of the coefficients 
$C_{\alpha i}$ and, therefore, of the elements of
the density matrix $\rho$. Therefore, the quantum circuit represented
by the diagram of states of Fig.~\ref{fig:dstates1qutrit} can be used 
to generate any given qutrit state $\rho$, once the ancillary qutrit
is disregarded.
Finally, we point out that the number of real parameters
$\{\alpha_i,\theta_k,\phi_l\}$ that determine the state $|\Psi\rangle$
is equal to $8$, that is, exactly to the number of parameters 
needed to determine a (generally mixed) single-qutrit state $\rho$.

It is clear from the purification drawn in Fig.~\ref{fig:dstates1qutrit} 
that we can write a generic single-qutrit state as 
\begin{equation}
\rho= \sum_{k=1}^3 p_k |\psi_k\rangle \langle \psi_k|,
\end{equation}
with 
$p_1=\cos^2\alpha_1\cos^2\alpha_2$, 
$p_2=\cos^2\alpha_1\sin^2\alpha_2$, 
$p_3=\sin^2\alpha_1$, 
$|\psi_1\rangle$ generic single-qutrit pure state,
$|\psi_2\rangle$ pure state residing in the two-dimensional 
subspace spanned by $|0\rangle$ and $|1\rangle$, and 
$|\psi_3\rangle=|0\rangle$. 
The picture developed in Sec.~\ref{sec:invasion}
about the ``invasion'' of the single-qubit Bloch ball by means of 
suitably scaled and translated pure-qubit Bloch spheres may be 
generalized to the single-qutrit case.
The role of the Bloch sphere is here played by the surface of
single-qutrit pure states and the volume of all single-qutrit
states is ``invaded'' when the parameters $p_k$ are varied, with 
the constraint $\sum_k p_k=1$.

\section{Two qubits}

\label{sec:2qubits}

We consider $M=N=4$ and set 
\begin{equation}
\left\{
\begin{array}{l}
C_{03}\in \mathbb{R}_+, \\
C_{12}\in \mathbb{R}_+, \,C_{13}=0,\\
C_{21}\in \mathbb{R}_+, \,C_{22}=C_{23}=0,\\
C_{30}\in \mathbb{R}_+, \,C_{31}=C_{32}=C_{33}=0.
%C_{03}\in \mathbb{R},\,C_{03}\ge 0, \\
%C_{12}\in \mathbb{R},\,C_{12}\ge 0, \,C_{13}=0,\\
%C_{21}\in \mathbb{R},\,C_{21}\ge 0, \,C_{22}=C_{23}=0,\\
%C_{30}\in \mathbb{R},\,C_{30}\ge 0, \,C_{31}=C_{32}=C_{33}=0.
\end{array}
\right.
\label{eq:gauge2qubits}
\end{equation}

We first obtain $C_{03}$ from 
\begin{equation}
\rho_{33}=|C_{03}|^2+|C_{13}|^2+|C_{23}|^2+|C_{33}|^2=|C_{03}|^2,
\end{equation}
then $C_{02}$, $C_{01}$, and $C_{00}$ from 
\begin{equation}
\rho_{23}=C_{02}C_{03}^\star+C_{12}C_{13}^\star
+C_{22}C_{23}^\star+C_{32}C_{33}^\star=
C_{02}C_{03},
\end{equation}
\begin{equation}
\rho_{13}=C_{01}C_{03}^\star+C_{11}C_{13}^\star
+C_{21}C_{23}^\star+C_{31}C_{33}^\star=
C_{01}C_{03},
\end{equation}
\begin{equation}
\rho_{03}=C_{00}C_{03}^\star+C_{10}C_{13}^\star
+C_{20}C_{23}^\star+C_{30}C_{33}^\star=
C_{00}C_{03},
\end{equation}
then $C_{12}$ from 
\begin{equation}
\rho_{22}=|C_{02}|^2+|C_{12}|^2+|C_{22}|^2+|C_{32}|^2=
|C_{02}|^2+C_{12}^2,
\end{equation}
then $C_{11}$ and $C_{10}$ from 
\begin{eqnarray}
\displaystyle
\rho_{12}=C_{01}C_{02}^\star+C_{11}C_{12}^\star
+C_{21}C_{22}^\star+C_{31}C_{32}^\star
\nonumber\\
\displaystyle=
C_{01}C_{02}^\star+C_{11}C_{12},
\end{eqnarray}
\begin{eqnarray}
\displaystyle
\rho_{02}=C_{00}C_{02}^\star+C_{10}C_{12}^\star
+C_{20}C_{22}^\star+C_{30}C_{32}^\star\nonumber\\
\displaystyle=
C_{00}C_{02}^\star+C_{10}C_{12}.
\end{eqnarray}
then $C_{21}$ from 
\begin{eqnarray}
\displaystyle
\rho_{11}=|C_{01}|^2+|C_{11}|^2+|C_{21}|^2+|C_{31}|^2
\nonumber\\
\displaystyle=
|C_{01}|^2+|C_{11}|^2+C_{21}^2,
\end{eqnarray}
then $C_{20}$ from
\begin{eqnarray}
\displaystyle
\rho_{01}=C_{00}C_{01}^\star+C_{10}C_{11}^\star
+C_{20}C_{21}^\star+C_{30}C_{31}^\star\nonumber\\
\displaystyle=
C_{00}C_{01}^\star+C_{10}C_{11}^\star+C_{20}C_{21},
\end{eqnarray}
and, finally, $C_{30}$ from 
\begin{equation}
\rho_{00}=|C_{00}|^2+|C_{10}|^2+|C_{20}|^2+C_{30}^2.
\end{equation}

\begin{figure}
\centerline{\epsfxsize=8.cm\epsffile{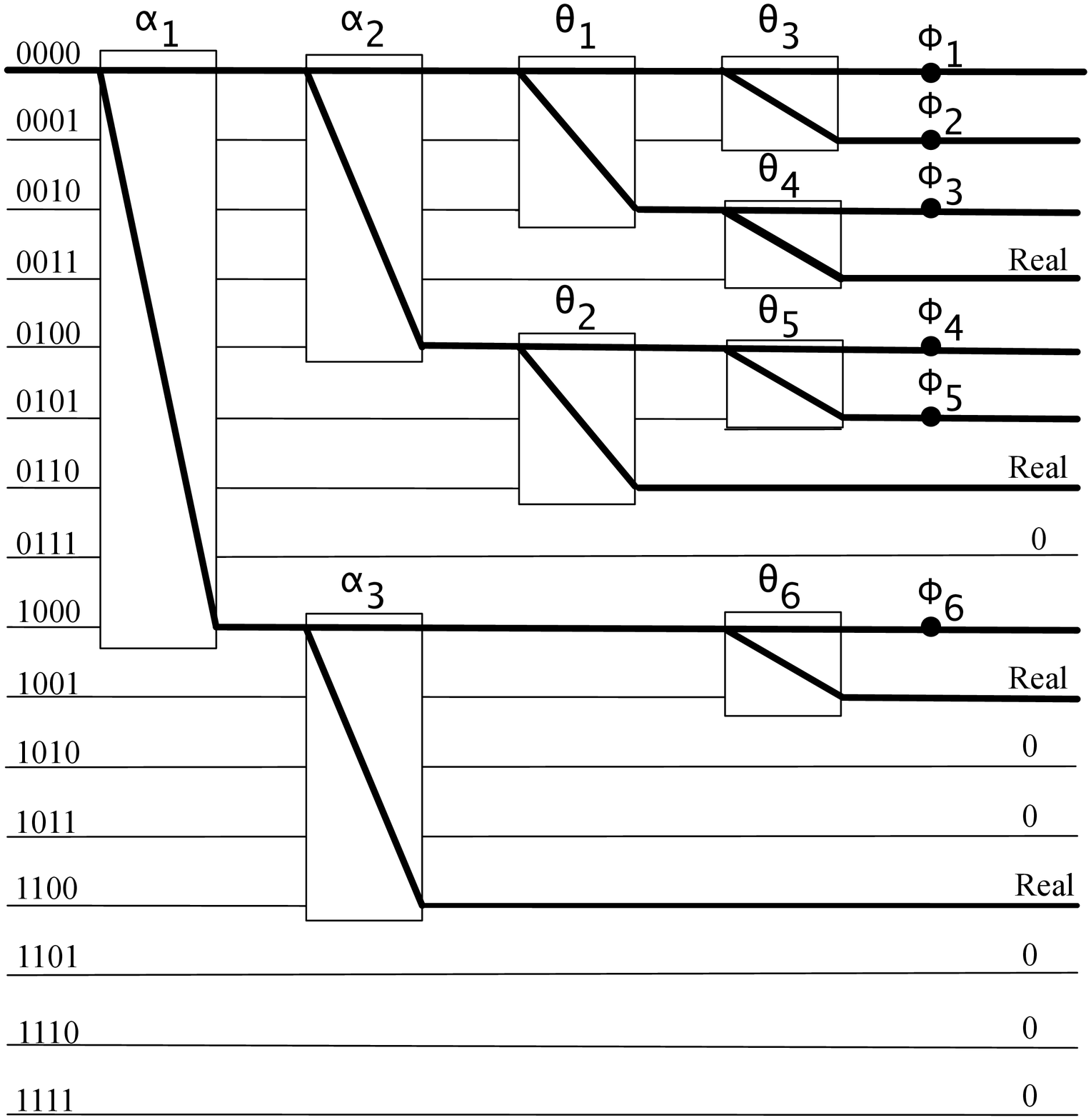}}
\caption{Diagram of states for the purification of a two-qubit state.
To simplify the plot, only the thick lines corresponding to the
information flow are shown inside the boxes.
The angles 
$\alpha_i,\theta_j\in\left[0,\frac{\pi}{2}\right]$, while the 
phases $\phi_k\in[0,2\pi)$.}
\label{fig:dstates2qubits}
\end{figure}

As for the previous examples, we point out that 
the number $N^2-1=15$ of free parameters determining the purification
state $|\Psi\rangle$ cannot be reduced given a generic 
two-qubit mixed state $\rho$. Moreover, conditions
(\ref{eq:gauge2qubits}) are determined from the 
diagram of states for the purification of a generic
two-qubit state shown in Fig.~\ref{fig:dstates2qubits}.
Following the information flow from the input states $|0000\rangle$
we can immediately write down the output purification state
$|\Psi\rangle=\sum_{\alpha,i} C_{\alpha i}|\alpha i\rangle$.
We obtain
\begin{equation}
\left\{
\begin{array}{l}
\displaystyle
C_{0000}=\cos\alpha_1 \cos\alpha_2 \cos\theta_1 \cos \theta_3 e^{i\phi_1},\\
\displaystyle
C_{0001}=\cos\alpha_1 \cos\alpha_2 \cos\theta_1 \sin \theta_3 e^{i\phi_2},\\
\displaystyle
C_{0010}=\cos\alpha_1 \cos\alpha_2 \sin\theta_1 \cos \theta_4 e^{i\phi_3},\\
\displaystyle
C_{0011}=\cos\alpha_1 \cos\alpha_2 \sin\theta_1 \sin \theta_4,\\
\displaystyle
C_{0100}=\cos\alpha_1 \sin\alpha_2 \cos\theta_2 \cos \theta_5 e^{i\phi_4},\\
\displaystyle
C_{0101}=\cos\alpha_1 \sin\alpha_2 \cos\theta_2 \sin \theta_5 e^{i\phi_5},\\
\displaystyle
C_{0110}=\cos\alpha_1 \sin\alpha_2 \sin\theta_2,\\
\displaystyle
C_{0111}=0,\\
\displaystyle
C_{1000}=\sin\alpha_1 \cos\alpha_3 \cos\theta_6 e^{i\phi_6},\\
\displaystyle
C_{1001}=\sin\alpha_1 \cos\alpha_3 \sin\theta_6,\\
\displaystyle
C_{1010}=0,\\
\displaystyle
C_{1011}=0,\\
\displaystyle
C_{1100}=\sin \alpha_1 \sin \alpha_3,\\
\displaystyle
C_{1101}=0,\\
\displaystyle
C_{1110}=0,\\
\displaystyle
C_{1111}=0.
\end{array}
\right.
\end{equation}
As in the previous cases, 
we can invert these equations and determine the parameters 
$\{\alpha_j,\theta_k,\phi_l\}$ in terms of the coefficients 
$C_{\alpha i}$ and, therefore, of the elements of
the density matrix $\rho$. 

We can see from Fig.~\ref{fig:dstates2qubits} 
that a generic two-qubit state can be written as
\begin{equation}
\rho= \sum_{k=1}^4 p_k |\psi_k\rangle \langle \psi_k|,
\end{equation}
with 
$p_1=\cos^2\alpha_1\cos^2\alpha_2$, 
$p_2=\cos^2\alpha_1\sin^2\alpha_2$, 
$p_3=\sin^2\alpha_1\cos^2\alpha_3$, and
$p_4=\sin^2\alpha_1\sin^2\alpha_3$.
The ``invasion'' picture developed in Sec.~\ref{sec:invasion}
can be extended also to the present case, with the volume of all
two-qubit states ``invaded'' when the parameters $p_k$ are varied,
under the constraint $\sum_k p_k=1$.
Note that the number of real parameters 
$\{\alpha_i,\theta_k,\phi_l\}$ used to determine the state 
$|\Psi\rangle$ is 15, that is, exactly the number of parameters 
required to determine a generic two-qubit state.

\section{$n$ qudits}

\label{sec:nqudits}

We consider $M=N=d^n$ and set 
\begin{equation}
\left\{
\begin{array}{l}
C_{0,N-1}\in \mathbb{R}_+, \\
C_{1,N-2}\in \mathbb{R}_+, \,C_{1,N-1}=0\\
C_{2,N-3}\in \mathbb{R}_+, \,C_{2,N-2}=C_{2,N-1}=0,\\
\vdots\\
C_{N-1,0}\in \mathbb{R}_+, \, C_{N-1,1}= 
C_{N-1,2}=\ldots \\
\quad =C_{N-1,N-1}=0.
%C_{0,N-1}\in \mathbb{R},\,C_{0,N-1}\ge 0, \\
%C_{1,N-2}\in \mathbb{R},\,C_{1,N-2}\ge 0,\,C_{1,N-1}=0\\
%C_{2,N-3}\in \mathbb{R},\,C_{2,N-3}\ge 0,\,C_{2,N-2}=C_{2,N-1}=0,\\
%\vdots\\
%C_{N-1,0}\in \mathbb{R},\,C_{N-1,0}\ge 0,\, C_{N-1,1}= 
%C_{N-1,2}=\ldots \\
%\quad =C_{N-1,N-1}=0.
\end{array}
\right.
\end{equation}

The general procedure for determining the coefficient 
$C_{\alpha i}$ is 
clear from the previous examples. 
\begin{itemize}
\item
We first determine 
$C_{0,N-1}$ from $\rho_{N-1,N-1}$, 
\item 
then 
$C_{0j}$ from $\rho_{j,N-1}$, with $j=N-2,...,0$, 
\item
then 
$C_{1,N-2}$ from $\rho_{N-2,N-2}$, 
\item
then 
$C_{1j}$ from $\rho_{j,N-2}$, with $j=N-3,...,0$, ..., 
\item
and, finally,
$C_{N-1,0}$ from $\rho_{00}$.
\end{itemize}

\begin{figure}
\centerline{\epsfxsize=8.cm\epsffile{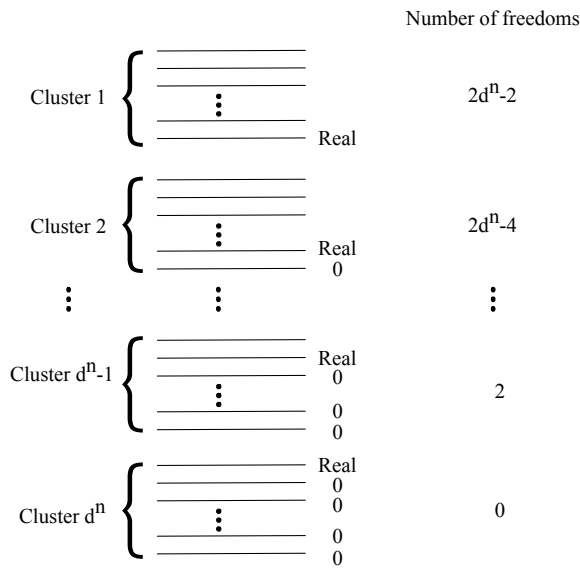}}
\caption{Schetch of the diagram of states for the purification 
of a $n$-qudit state by means of $2n$ qudits.
The diagram has $d^{2n}$ lines, one for each state of the
computational basis. 
We can cluster groups of
$d^n$ lines. The constraints and the number 
of real freedoms for each cluster are highlighted.
If we add these numbers plus the $d^n-1$ independent 
weights in the mixture (\ref{eq:pknqudits}), we obtain a total 
number of $d^{2n}-1$ degrees of freedom.}
\label{fig:dstatesnqudits}
\end{figure}

This purification is optimal as the number of qudits 
of the ancillary system cannot be reduced. 
The number of real free parameters that must be set to determine 
a density matrix of size $d^n$ is $d^{2n}-1$ [the $-1$ term comes from the 
normalization condition ${\rm Tr}(\rho)=1$].  
Therefore, an ancillary system of $n-1$ qudits is not sufficient
to purify $\rho$, as a pure state $|\Psi\rangle$
in the Hilbert space of $n+(n-1)=2n-1$ qudits
has $2d^{2n-1}-2$ freedoms (the $-2$ term is due to the normalization 
condition for $|\Psi\rangle$ and to the fact that a global phase factor
in $|\Psi\rangle$ is arbitrary). This number is not sufficient as 
$2d^{2n-1}-2<d^{2n}-1$ for any $d\ge 2$, $n\ge 1$.
On the other hand, as illustrated Fig.~\ref{fig:dstatesnqudits}, 
the number of free real parameters in our purification 
method is exactly $d^{2n}-1$. 
From the schematic drawing in Fig.~\ref{fig:dstates2qubits} we
can also see that the $n$-qudit state can be written as
\begin{equation}
\rho=\sum_{k=1}^{d^n} p_k |\psi_k\rangle\langle \psi_k|,
\label{eq:pknqudits}
\end{equation}
with the pure states from $|\psi_1\rangle$ to $|\psi_{d^n}\rangle$
residing in subspaces of decreasing dimension, from $d^n$ to $1$.

We note that the tensor product structure of many-qudit 
quantum systems does not play any role in our purification 
protocol. That is to say, the same purification scheme applies
to a system of $n$ qudits or to a single system of size $N=d^n$.
Of course, the practical implementation of the protocol will
depend on the specific quantum hardware at disposal.

From the mathematical viewpoint, our purification protocol 
can be seen as the Cholesky 
decomposition~\cite{golub} of the density matrix $\rho$.
If we consider the coefficients $C_{\alpha i}$ as elements of a 
$d^n \times d^n$ matrix $C$, then its transpose $D\equiv C^T$ is 
a upper triangular matrix and 
Eq.~(\ref{purecondition}) reads 
\begin{equation}
\rho = C^T C^\star = D D^{T*} = D D^\dagger,
\end{equation}
namely it is the Cholesky decomposition of 
the density matrix $\rho$. Such decomposition is unique 
when $\rho$ is positive, while the ambiguities arising when 
$\rho$ is singular may be removed by suitable prescriptions 
or reshuffling as previously discussed for the single-qubit case. 

It is interesting to remark that the Cholesky decomposition 
has already been used in the context of quantum information science
in parametrizing the density operator in order to 
guarantee positivity~\cite{dariano}. The purpose of Ref.~\cite{dariano}
was to present a universal technique for quantum-state estimation.

\section{Final remarks}

\label{sec:conclusions}

In summary, we have proposed an algorithm for the purification 
of a generic $n$-qudit state. This algorithm is optimal, in that
the number $n$ of ancillary qudits used for the purification cannot
be further reduced. Moreover, our algorithm can also be seen
as a quantum protocol for the generation of a generic $n$-qudit
state by means of suitable unitary operations applied both to
the system and to the ancillary qudits, with the ancillary qudits
eventually disregarded.
While also the well-known purification
(\ref{eq:spectralpur}) based on the spectral decomposition 
(\ref{dmatdiag}) uses $n$ ancillary qudits, our method 
is optimal in that it readily provides a 
purification state that depends on a number of parameters 
exactly equal to the number of degrees of freedom of the generic 
$n$-qubit state that we wish to purify.

It is well known that the purification $|\Psi\rangle$ of a 
generic mixed state $\rho$ cannot be uniquely determined, 
as the partial trace over the ancillary qudits is invariant 
under any unitary transformation 
$U_A\otimes \openone_S$ acting non trivially on 
the ancillary qudits only. Indeed, we have
\begin{equation}
\rho={\rm Tr}_A (|\Psi\rangle\langle\Psi|)=
{\rm Tr}_A (|\Psi'\rangle\langle\Psi'|),
\end{equation} 
with 
\begin{equation}
|\Psi'\rangle=(U_A\otimes \openone_S)|\Psi\rangle.
\end{equation}
It is sufficient to add the unitary transformation 
$U_A\otimes \openone_S$ at the end of our purification protocol 
to obtain any $2n$-qudit purification $|\Psi'\rangle$ 
of a $n$-qudit state $\rho$. On the other hand,
we can say that our protocol selects a very convenient 
purification as the coefficients of the wave function 
$|\Psi\rangle$ in the computational basis are easily 
determined from the density matrix $\rho$ and the quantum 
circuit generating $\rho$ can be immediately drawn. 

\appendix

\section{Diagrams of states}

\label{app:dstates}

Diagrams of states~\cite{diagrams} graphically represent how quantum
information is elaborated during the execution of a quantum circuit.
In the usual way of drawing a quantum 
circuit~\cite{qcbook,nielsen} each horizontal line represents a qubit. 
In contrast, in diagrams of states we draw a horizontal line for each
state of the computational basis. Therefore, diagrams of states are less 
synthetic but may help us to clearly visualize quantum information
flow in a quantum circuit. 

\begin{figure}
\centerline{\epsfxsize=8.cm\epsffile{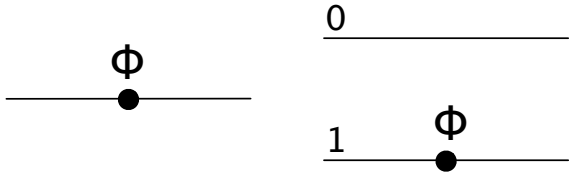}}
\caption{Quantum circuit (left) and diagram of states (right) for 
the phase-shift gate.}
\label{fig:phasegate}
\end{figure}

\begin{figure}
\centerline{\epsfxsize=8.cm\epsffile{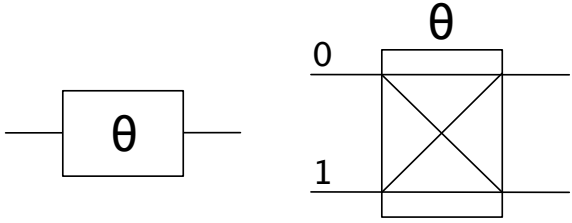}}
\caption{Quantum circuit (left) and diagram of states (right) for 
the $R(\theta)$ gate.}
\label{fig:thetagate}
\end{figure}

\begin{figure}
\centerline{\epsfxsize=8.cm\epsffile{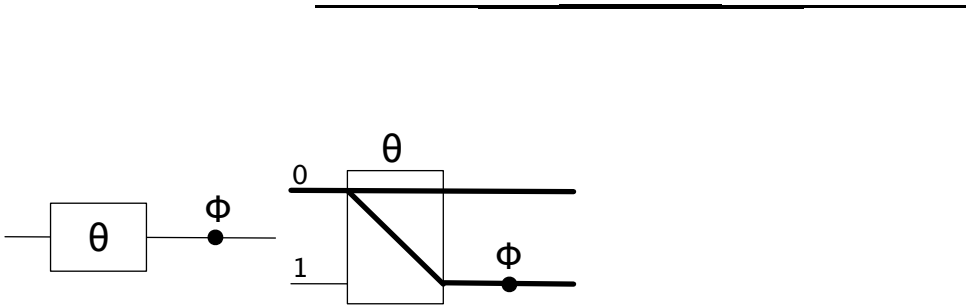}}
\caption{Quantum circuit (left) and diagram of states (right) for 
the generation of a generic single-qubit state.
To simplify the plot, only the thick lines corresponding to the
information flow are shown inside the boxes.}
\label{fig:1qgeneration}
\end{figure}

For the purposes of the present paper, it will be sufficient to 
show the diagrams of states for elementary single-qubit quantum gates.
The phase-shift gate 
${\rm PHASE}(\phi)$, defined by Eq.~(\ref{eq:PHASE}),
and the rotation gate 
$R(\theta)$, defined by Eq.~(\ref{eq:Ralpha}), are shown 
in Fig.~\ref{fig:phasegate} and Fig.~\ref{fig:thetagate}, respectively.
Finally, the generation of a generic single-qubit state 
%$\cos\frac{\theta}{2} |0\rangle+\sin\frac{\theta}{2}e^{i\phi}|1\rangle$
$\cos{\theta} |0\rangle+\sin{\theta} e^{i\phi}|1\rangle$
starting from the input 
state $|0\rangle$ is shown in Fig.~\ref{fig:1qgeneration}.
The information flows on the thick lines, from left to right,
while thinner lines correspond to absence of information. 
Note that, following the thick lines, the final state can here be 
immediately written.

\end{document}